\documentclass[twocolumn,showpacs,amsmath,amssymb]{revtex4}
\usepackage{graphicx}
\usepackage{dcolumn}
\usepackage{bm}

\newcommand{\n}[1]{\label{#1}}

\newcommand{\ba}{\begin{eqnarray}}
\newcommand{\ea}{\end{eqnarray}}
\newcommand{\be}{\begin{equation}}
\newcommand{\ee}{\end{equation}}
\newcommand{\bd}{\begin{displaymath}}
\newcommand{\ed}{\end{displaymath}}

\newcommand{\lap}{\bigtriangleup}
\newcommand{\Rot}{\mbox{curl}}
\newcommand{\Div}{\mbox{div}}
\newcommand{\pa}{\partial}

\newcommand{\h}[1]{(#1)}
\newcommand{\z}{\zeta}
\newcommand{\bz}{\bar{\zeta}}
\newcommand{\w}[1]{\omega^{(#1)}}
\newcommand{\bw}[1]{\bar{\omega}^{(#1)}}

\newcommand{\hh}{\, ,\hspace{0.5cm}}

\begin{document}

\draft


\title{Gravitational field of a spinning radiation beam-pulse in
higher dimensions}
\author{Valeri P. Frolov$^*$ and Dmitri V. Fursaev$^\dagger$}
\address{
  \medskip
  $^*$Theoretical Physics Institute, 
  Department of Physics, University of Alberta\\
  Edmonton, AB, Canada, T6G 2J1\\
  {\rm E-mail: \texttt{frolov@phys.ualberta.ca}}\\
  \medskip
}
\address{
  $^\dagger$ Dubna International University and\\
  Bgoliubov Laboratory of Theoretical Physics\\
  Joint Institute for Nuclear Research\\
  141 980, Dubna, Moscow Region, Russia\\
  {\rm E-mail: \texttt{fursaev@thsun1.jinr.ru}}\\
  \medskip
}
\date{\today}

\begin{abstract}
We study the gravitational field of a spinning
radiation beam-pulse in a higher dimensional spacetime. We derive
first the stress-energy tensor for such a beam in a flat spacetime
and find the gravitational field generated by it in the linear
approximation. We demonstrate that this gravitational field can also
be obtained by boosting the Lense-Thirring metric in the limit when
the velocity of the boosted source is close to the velocity of light.
We then find an exact solution of the Einstein equations describing
the gravitational field of a polarized radiation beam-pulse in a
space-time with arbitrary number of dimensions. In a $D-$dimensional
spacetime this solution contains $[D/2]$ arbitrary functions of one
variable (retarded time $u$), where $[d]$ is the integer part of $d$.
For the special case of a 4-dimensional spacetime we study effects
produced by such a relativistic  spinning beam on the motion of test
particles and light.
\end{abstract}

\pacs{04.70.Bw, 04.50.+h, 04.20.Jb \hfill Alberta-Thy-02-05}

\maketitle

\section{Introduction}

A metric for the gravitational field of a massive source moving
with the velocity close to the velocity of light is known for many
years. This metric was discovered by Aichelburg-Sexl in 1971
\cite{AiSe}. It can be obtained by boosting the Schwarzschild metric.
In the so-called  Penrose limit, when the boost becomes infinite
while the energy remains finite, the boosted Schwarzschild metric
takes the form of a gravitational shock-wave. The properties of such
solutions and their generalizations, as well as related
references can be found in a recent book \cite{BaHo}.

There are several reasons why such boosted solutions are of interest
and can be helpful. In recent years the Penrose limit attracted a lot
of attention in the string theory after demonstration that the theory can
be exactly solvable in this limit \cite{string}. In the type IIB
supergravity a certain gravitational plane-wave solution constitutes
a maximally supersymmetric background for the IIB string. In the
light-cone gauge the string theory $\sigma$-model on the plane-wave
background is reduced to a free massive two-dimensional model. This
model, similarly to what happens in a flat background, is solvable
and can be easily quantized. Moreover, there exists a duality relating
type IIB superstrings in the maximally supersymmetric plane-wave
background to a four-dimensional ${\cal N}=4$ super Yang-Mills theory.

Another application is related to studying mini black hole formation
in a high-energy collision of two particles (for general discussion,
see e.g a review \cite{Kant}).  This problem was motivated by
theories with large extra dimensions. For two highly relativistic
particles the gravitational field of each of them, as observed in the
center of mass frame, can be approximated by the Aichelburg-Sexl
metric. Eardley and Giddings \cite{EaGi:02} used this approximation
to estimate the cross section for black hole production. This paper
also discusses black hole formation in the presence of extra
dimensions. Generalization of these results for higher dimensional
head-on and not head-on collisions can be found in
\cite{YoNa:02,YoNa:03,YoRy:05}. The paper \cite{YoNa:02} discusses the
formation of an apparent horizon for the head-on collision from the
point of view of the hoop conjecture. A discussion of the higher
dimensional generalizations of the hoop conjecture can be found in
\cite{BaFrLe:04}.

If a black hole is created in the non head-on collision of two
ultrarelativistic particles, one can expect that it would be rapidly
rotating and the angular momentum of the black hole will be important
for its further evolution \cite{EaGi:02,FrSt:02a,FrSt:02b,IdOdPa,JuKiPa}.
The cross section for the production of a rotating higher
dimensional black hole and  a black ring  was estimated in
\cite{IdOdPa}.

If colliding ultrarelativistic particles have spin, one can expect
new effects connected with a spin--orbit and/or spin--spin
interaction. In order to study these effects one needs to know the
shock wave solutions describing the gravitational field of particles
with spin. A natural way to obtain such solutions is to boost the
Kerr metric. The lightlike boost of the Kerr black hole in the
direction parallel to its spin was discussed in
\cite{FePe:90,LoSa:92}. It was shown later that the method used in
the paper \cite{LoSa:92} contains an ambiguity. This ambiguity was
fixed in more recent publications. By using different approaches a
solution for the boosted Kerr black holes was obtained and studied in
\cite{BaNa:95,BaNa:96,BuMa:00, BaHo:03,BaHo:04}. The papers
\cite{BuMa:00,BaHo:03,BaHo:04} discuss also a general case when direction
of the boost differs from the spin direction.  Relativistic boosted
metrics for higher dimensional rotating black holes were obtained in
\cite{Yosh:05}.

In the derivations of the lightlike boosted Kerr metric it is usually
assumed that a rotation parameter $a$ is fixed, while the mass $M$
tends to zero, in order to keep fixed the energy, $E=\gamma M$. Here
$\gamma=(1-v^2/c^2)^{-1/2}$ and $v$ is the velocity of the black hole.
Thus in the Penrose limit the angular momentum $J=aM$ also vanishes.
This means that at a finite distance from the black hole the field
(curvature) becomes weak in the comoving reference frame. For this
reason in order to obtain a boosted Kerr metric in the Penrose limit
it is sufficient to start with a Lense-Thirring weak field solution.

For a discussion of  black hole formation in the relativistic
collision of particles with a spin, it is important to study another
limit when their internal angular momentum (spin) does not vanish
when $\gamma\to\infty$. The aim of this paper is to obtain such
solutions and describe their properties. We start by considering a
narrow pulse-like beams of electromagnetic radiation (section~2).  We
assume that this radiation is almost monochromatic and is either
right or left circularly polarized. We describe this beam in the
geometric optics approximation assuming that the duration of the beam
$L$ and the radius of its cross-section $\varrho$ are much larger than the
wavelength $\lambda$. Since our aim is  studying  the gravitational
field created by such sources  both in four and higher dimensions, we
perform the calculation of the stress-energy tensor
and the angular momentum
of the beam-pulse in flat spacetime with the number of dimensions $D\geq
4$.

In section~3 we derive the gravitational field of the  beam-pulse of
spinning radiation in the linear approximation. In appendix~A it is
demonstrated that the same metric can be obtained in the Penrose
limit from the Lense-Thirring metric, provided during the
limit-process the angular momentum $J$, as well as the energy $E$ are
fixed. In four-dimensional case such a boosted solution is
characterized by 2 parameters, $E$ and $J$. In the higher dimensional
case the number of angular momentum parameters $J_i$ is greater than
one. It coincides with a number $l$ of independent bi-planes of
rotation orthogonal to the direction of motion. For a $D-$dimensional
spacetime $l$ is the integer part of the ratio $(D-2)/2$.

For a fixed angular momentum  the solution at a finite distance from
the source does not reduce to its weak field limit. For this reason
in general case a boosted weak field solution is not sufficient for
the generation of the exact solution of the beam-pulse of spinning
radiation.  In the second part of the paper we derive such an exact
solution and study its properties. We present the exact solution in
section~4. A proof that the presented metric is Ricci-free outside
the source is given in appendix~B. The form of the metric differs
slightly for even and odd number of spacetime dimensions. The
obtained solution contains $l+1$ arbitrary functions of one variable
("retarded time" $u$) which determine the profiles of the
distributions of the energy and of the angular momenta of the
beam-pulse. The special case when $D=4$ is discussed in section~5. In
section~6 we discuss motion of particles and light in the
gravitational field of the beam-pulse of spinning radiation. We focus
our attention on the four-dimensional case and demonstrate that the
effect produced by the spin creates a force on a particle similar to
the usual centrifugal repulsive force,  while the energy produces the
attractive "Newtonian" force. We discuss possible applications of the
obtained results in "Summary and Discussions" (section~7).

\section{Beam-pulses of circularly polarized radiation}

\subsection{Beams of circularly polarized electromagnetic field
in a four-dimensional spacetime}

Let us discuss properties of the polarized radiation.
We consider first the electromagnetic field $A_{\mu}$ in a flat
4-dimensional spacetime. We choose the first two coordinates to be
null $x^1=u=t-\xi$, $x^2=v=t+\xi$, and denote the other spatial
coordinates $x^a$, $a=3,4$.
In a Lorentz gauge $(A^{\mu}_{\ ,\nu}=0)$ the
electromagnetic field obeys the equation
\be
\Box A_{\mu}=(-4\partial_{u}\partial_{v}+\partial_{\perp}^2) A_{\mu}=0\, .
\ee
Here $\partial_{\perp}^2=\partial_3^2+\partial_4^2$. A solution for a
circularly polarized  monochromatic plane-wave propagating in
$\xi-$direction has the form
\be\n{AA}
A_{\mu}=  {\cal A} \left[ e_{\mu} \exp(-i\omega u)
+\bar{e}_{\mu} \exp(i\omega u)\right]\, .
\ee
Here ${\cal A}$ is a real amplitude and $e_{\mu}$ is a complex
null vector ($\bar{e}_{\mu}$ is its complex conjugate) in
the bi-plane $(x^3,x^4)$, orthogonal to the direction of
wave propagation. One can choose $e_{\mu}=e^{+}_{\mu}$ or
$e_{\mu}=e^{-}_{\mu}$, where
\be
e^{\pm}_{\mu}={1\over \sqrt{2}}(\delta_{\mu}^3 \pm i\delta_{\mu}^4)\,
.
\ee
The superscripts $+$ and $-$ correspond to the right and left
circularly polarized waves, respectively.

If ${\cal A}$ is not a constant, but a slowly changing function,
(\ref{AA}) is an approximate solution provided  $\omega$ is large. If
$\nabla {\cal A}\sim \varrho^{-1} {\cal A}$ then (\ref{AA}) gives a
solution in the geometric optics approximation provided $\lambda/\varrho\ll
1$, where $\lambda=\omega^{-1}$ is a wave-length. We assume that
${\cal A}$ depends  on ${\bf x}_{\perp}=(x^3,x^4)$, is localized
in the vicinity of ${\bf x}_{\perp}=0$, and vanishes outside the
radius $R=\sqrt{{\bf x}_{\perp}^2}\sim \varrho$ \cite{Masl}.

The field strength $F_{\mu\nu}=A_{\mu,\nu}-A_{\nu,\mu}$ can be written
as
\be
F_{\mu\nu}={\cal F}_{\mu\nu}\exp(-i\omega u)+\bar{{\cal
F}}_{\mu\nu}\exp(i\omega u)\, ,
\ee
where the complex tensor ${\cal F}_{\mu\nu}$ has the following
non-vanishing components
\be
{\cal F}_{u a}=i\omega {\cal A} e_{a}\hh
{\cal F}_{a b}= e_{a}{\cal A}_{,b}-e_{b}{\cal A}_{,a} \, .
\ee

The metric stress-energy tensor of the electromagnetic field is
\be \n{3.3}
t^\mu_\nu=F^{\mu\lambda}F_{\nu\lambda}-\frac 14 \delta^\mu_\nu
F_{\alpha\beta}F^{\alpha\beta} ~~.
~~
\ee
The stress-energy tensor for a monochromatic wave besides a part
independent of $u$ contains also a rapidly oscillating contribution
$\sim \exp (\pm 2i\omega u)$. This contribution
vanishes after averaging over the time interval $L\gg \lambda$ and can
be neglected. Thus one has the following expression for the
averaged stress-energy tensor
\be
t^\mu_\nu=2\Re \left[{\cal F}^{\mu\lambda}\bar{\cal
F}_{\nu\lambda}
-\frac 14 \delta^\mu_\nu
{\cal F}_{\alpha\beta}\bar{\cal F}^{\alpha\beta}\right] ~~,
\ee
where $\Re [a]$ denotes a real part of $a$.

It is easy to check that the components ${t}_{vv}$ and  ${t}_{va}$ vanish
identically. The components of ${t}_{uv}$, ${t}_{ab}$ are
of the order of $\lambda^2 /l^2$ and also can be neglected.
The leading non-vanishing components of ${t}_{\mu\nu}$ are
\be \n{3.8}
{t}_{uu}=2\omega^2 {\cal A}^2~~~,
\ee
\be \n{3.9}
{t}_{ua}=i\omega {\cal A}{\cal A}_{,b}(\bar{e}_a e_b
-e_a\bar{e}_b)~~~.
\ee
The stress-energy tensor  (\ref{3.9}) obeys the relation ${t}_{ua,a}=0$.

The energy of the beam-pulse is defined as follows
\be \n{3.12}
E=\int du\,  d{\bf x}_{\perp} t_{uu}~~~.
\ee
Since the averaged stress-energy tensor $t_{uu}$ does not depend on
$u$, the energy $E$ of the beam is divergent. In order to deal with a
realistic system with finite energy it is sufficient to assume that
the beam-pulse has a finite duration $L\gg \lambda$ in time.  To deal
with this situation one can use the geometric optics approximation
(\ref{AA}) and allow ${\cal A}$ to depend (slowly) on $u$. We adopt
a simpler approach. Namely  we assume that during the time interval
$u\in (-L/2,L/2)$ the stess-energy tensor is given by
(\ref{3.8})-(\ref{3.9}), and vanishes outside this interval.
We denote by $T_{\mu\nu}$ the corresponding stress-energy tensor. It
can be written as
\be \n{2.6}
T_{\mu\nu}(u,{\bf x}_{\perp})=\chi(u) Lt_{\mu\nu}({\bf x}_{\perp})~~,
\ee
where
\be \n{3.7}
\chi(u)={1 \over L}(\vartheta(u+L/2)-\vartheta(u-L/2))~~~,
\ee
and $\vartheta(u)$ is the Heaviside step function.
Thus one has
\be \n{3.12a}
E=\int du\,  d{\bf x}_{\perp} T_{uu}=
L\int d{\bf x}_{\perp} t_{uu}~~~.
\ee
Using this relation we obtain
\be \n{3.10}
E=2NL\omega^2 ~~~,
\ee
where $N=\int d{\bf x}_{\perp} {\cal A}^2$ is a normalization constant
depending on the amplitude ${\cal A}$.

In the Minkowski space-time one can also
define the conserved angular momentum $J_{ab}$ of the system.
This can be done with the help of the
angular momentum tensor \cite{Noether}
\be \n{3.4}
M^{\sigma\rho}_{\ \ \nu}=x^\sigma T^\rho_{\ \nu}-x^\rho T^\sigma_{\
\nu}~~
\ee
as follows
\be \n{3.13}
J^{ab}=\int du d{\bf x}_{\perp} M_{\ \ \ u}^{ab}~~~.
\ee
Using (\ref{3.9}) one obtains
\be \n{3.15}
J_{ab}=i{E \over 2\omega} (e_a\bar{e}_b-\bar{e}_a e_b)~~~.
\ee

For simplicity we assume that the beam-pulse is axisymmetric.
In this case
the components of the spin tensor $J_{ua}$, $J_{va}$
vanish. The components $J_{uv}$ may be nontrivial but they are not
relevant for further analysis because their contribution to the
gravitational field (the $uv$ component of the metric)  is of
higher order with respect to the contribution produced by the
energy of the beam.

In a four-dimensional spacetime the tensor $J_{ab}$ can be used to
define the vector of  spin. Denote by ${\bf n}$ a unit vector in
the direction of the wave propagation.
Then the vector of the spin is $J {\bf n}$, where
\be \n{3.15a}
J=\varepsilon_{ab} J^{ab}=i{E \over \omega}
\varepsilon_{ab}e_a\bar{e}_b~~~
\ee
This vector is directed along the beam axis and its value is
\be
J=\pm {E\over \omega}\, ,
\ee
for the right ($+$) and left ($-$) polarization, respectively.

For a monochromatic wave which is not a state with a given helicity,
but is a superposition of the right and left polarized radiation, the
energy $E$ and the
total angular momentum $J$ are
\be\n{EJ}
E=\omega(J^+ +J^-)\hh J=J^+ -J^-\, .
\ee
Similar calculations with similar conclusions can be easily done for
a high frequency monochromatic beam-pulse of gravitational
radiation.

\subsection{Higher dimensional case}

Consider now the higher dimensional generalization of the results obtained
in the previous section. We again use the coordinates $u=t-\xi$ and
$v=t+\xi$ as the first two coordinates, $x^1=u$ and $x^2=v$, and
denote the other spatial coordinates by $x^a$, $a=3,\ldots, D$, where $D$
is the number of spacetime dimensions. We shall also use a vector
notation ${\bf x}_{\perp}=(x^3,\ldots,x^D)$.
The vector potential
of an "axially symmetric" monochromatic beam  can be written as
\be \n{3.1-a}
A_u=A_v=0~~,
\ee
\be \n{3.2-a}
{\bf A}(u,R)=\sum_{s=3}^{D}\breve{\bf e}_{s} \left[b_s(R)e^{-i\omega u}
+\bar{b}_s(R) e^{i\omega u}
\right]~~.
\ee
Here ${\bf A}$ denotes a vector with components $A_a$. $\breve{\bf
e}_{s}$  are real polarization vectors normalized as $(\breve{\bf
e}_{s},\breve{\bf e}_{s'})=\delta_{ss'}$. The coefficients $b_s$  are complex functions of
${\bf x}_{\perp}$, $\bar{b}_s$ are their complex conjugates.
As earlier we consider a beam of radiation and assume
that these functions vanish at $R \ge \varrho$, where $R=\sqrt{{\bf
x}_{\perp}^2}$.

Computation of the energy and the angular momentum of the beam is
analogous to the calculations performed in the previous section. It
yields
\be \n{10.1}
E=2L\omega^2\int d{\bf x}_{\perp} \sum_{s}b_s\bar{b}_s~~~,
\ee
\be \n{10.2}
J^{ab}=\sum_{ss'}K_{ss'}\breve{e}_{s}^{\ a}\breve{e}_{s'}^{\ b}~~~,
\ee
\be \n{10.3}
K_{ss'}=iL\omega\int d{\bf x}_{\perp}
(\bar{b}_sb_{s'}-b_{s}\bar{b}_{s'})~~~.
\ee
As earlier $L$ is the duration of the beam-pulse.

The matrix $K_{ss'}$ is a real antisymmetric $(D-2)\times (D-2)$ matrix.
There exists an orthogonal transformation $O$ which brings it to a
block-diagonal form where its only non-vanishing components (for a
suitably chosen numeration) are $K_{34}=-K_{43}$,
$K_{56}=-K_{65}$ etc. (see \cite{Gant}).
If $D$ is odd the last row and column of the above matrix vanish.

The transformation $O$ obeys the property $O^TO=I$, where $T$ is a
transposition, and $I$ denotes a unit matrix. The matrix $O$
transforms an old orthogonal basis  $\breve{\bf e}_{s}$ into a new
one, $\breve{\bf e}'_{s'}=O_{s'}^{~~s}\breve{\bf e}_{s}$. In the new
basis, where $K_{ss'}$ has the block-diagonal form, pairs of vectors
$\{\breve{\bf e}_{3},\breve{\bf e}_{4}\}$, $\{\breve{\bf
e}_{5},\breve{\bf e}_{6}\}$ and etc. define so-called bi-planes of
rotation. The number of these rotation planes is $l=[(D-2)/2]$, where
$[a]$ is an integer part of $a$. We enumerate the rotation
planes by an index $i=1,\ldots,l$.
We denote by ${\bf e}_i$ a complex null vector spanning the {\it i}-th
bi-plane,
\be
\n{10.4} {\bf e}_i={1 \over \sqrt{2}}(\breve{\bf e}_{2i+1}+i\breve{\bf
e}_{2i+2})~~.
\ee
These complex vectors are normalized as
\be
({\bf e}_i, {\bf e}_j)=(\bar{\bf e}_i, \bar{\bf e}_j)=0\hh ({\bf e}_i,
\bar{\bf e}_j)=\delta_{ij}\, .
\ee

If $D$ is even, the transverse part (\ref{3.2-a})
of the vector potential can be rewritten in the form
\be \n{10.5}
{\bf A}=\sum_{i=1}^{(D-2)/2}\left[(a_i{\bf e}_i+\bar{c}_i\bar{\bf
e}_i)e^{-i\omega u}+ (\bar{a}_i\bar{\bf e}_i+c_i{\bf e_i}) e^{i\omega
u} \right]~~,
\ee
where $a_i, c_i$ are complex functions which are
expressed in terms of $b_s$ as
\be \n{10.6}
a_i={1 \over
\sqrt{2}}(b_{2i+1}-ib_{2i+2})~~,~~ c_i={1 \over
\sqrt{2}}(b_{2i+1}+ib_{2i+2})~~.
\ee
If $D$ is odd, there exists an extra
term in (\ref{10.5}) corresponding to the contribution of a (real)
polarization vector ${\bf e}_0$ orthogonal to all the rotation bi-planes.
We denote by  $b_0$ the corresponding amplitude coefficient.

Let us define the angular momentum $J_i$ in the $i$-th bi-plane as
$J_i=2K_{(2i+1)(2i+2)}$. Then  by using (\ref{10.1})-(\ref{10.3}) it
can be shown that the following relations hold
\be \n{10.7}
E=\omega\sum_i(J_i^+ +J_i^-)+E_0~~,
\ee
\be \n{10.8}
J_i=J_i^+ -J_i^-~~,
\ee
\be \n{10.9}
J_i^+ =\omega L\int d{\bf x}_{\perp}\,  c_i\bar{c}_i~~,~~J_i^-
=\omega L\int d{\bf x}_{\perp} \, a_i\bar{a}_i~~,
\ee
\be \n{10.10}
E_0 =2L\omega^2\int d{\bf x}_{\perp} \, b_0\bar{b}_0~~.
\ee
$E_0=0$ if $D$ is even. Formula (\ref{10.7}) is a generalization of
the relation (\ref{EJ}) between the energy of the beam and its
spin. Quantities $J_i^\pm$ correspond to "left" and "right"
polarizations in the $i$-th bi-plane.

\section{Gravitational field of a beam-pulse in the weak field approximation}

We look for the gravitational field of a source which moves with the
velocity of light in a flat $D$-dimensional space time. The source is
a beam-pulse of circularly polarized radiation. For brevity we call
such a source `gyraton'. The gyraton is supposed to be stretched along the axis
of the motion and it moves rigidly, i.e., it preserves its profiles
both in the longitudinal and transverse directions. As earlier we use
coordinates $x^\mu=(u,v,x^a)$, where $x^a$, $a=3,..,D$, are the
transverse coordinates, and ${\bf x}_{\perp}$ is a vector in the
transverse direction. The gyraton is moving
along the $\xi$-axis in the positive direction. We assume that only non-vanishing
components of the stress-energy
tensor of the gyraton are $T_{uu}(u,{\bf x}_{\perp})$ and $T_{ua}(u,{\bf x}_{\perp})$.
This tensor is divergence free if $\partial^a T_{ua}(u,{\bf x}_{\perp})=0$.

Let $\eta_{\mu\nu}$ be the Minkowski metric
\be
ds_0^2=\eta_{\mu\nu}dx^{\mu} dx^{\nu}=-du\, dv +d{\bf x}_{\perp}^2
\ee
In the linear approximation the gravitational field is
$g_{\mu\nu}=\eta_{\mu\nu}+h_{\mu\nu}$, where the field perturbations
$h_{\mu\nu}$ obey the equation
\be \n{2.1}
\Box \tilde{h}_{\mu\nu}=-\kappa T_{\mu\nu}~~,
\ee
\be \n{2.1a}
\tilde{h}_{\mu\nu}=h_{\mu\nu}-\frac 12\eta_{\mu\nu} h~~.
\ee
Here $\kappa=16\pi G$, $G$ is the higher dimensional gravitational
coupling constant, and
\be
\Box=-4\partial_{u}\partial_{v}+\Delta\hh \Delta=\partial_{\perp}^2\, .
\ee

For a beam-pulse of radiation $T_{\mu}^{\mu}=0$. Thus, in the equation
(\ref{2.1}) instead of $\tilde{h}_{\mu\nu}$ one can put
${h}_{\mu\nu}$. Notice also that if ${h}_{\mu\nu}({\bf x}_{\perp})$
is a solution for $T_{\mu\nu}({\bf x}_{\perp})$ then for an arbitrary
function $\chi(u)$, function $\chi(u){h}_{\mu\nu}({\bf x}_{\perp})$ is a solution
for a source $\chi(u) T_{\mu\nu}({\bf x}_{\perp})$. We shall use this
property to construct solutions for a source with time $u$
dependent profiles. For the moment we consider a source
$T_{\mu\nu}=t_{\mu\nu}({\bf x}_{\perp})$ which does not depend on
$u$. To obtain $h_{\mu\nu}$ one needs to solve the equation
\be\n{DEL}
\Delta h_{\mu\nu}=-\kappa t_{\mu\nu}({\bf x}_{\perp})\, .
\ee

Denote by ${\cal G}({\bf x}_{\perp},{{\bf x}'}_{\perp})$ the Green's
function for the operator $\Delta$ in the transverse space
\be \n{G1}
\Delta {\cal G}({\bf x}_{\perp},{{\bf x}'}_{\perp})
=\delta ({\bf x}_{\perp}-{{\bf x}'}_{\perp})\, .
\ee
Denote $\rho=|{\bf x}_{\perp}-{{\bf x}'}_{\perp}|$ then
\be \n{G2}
{\cal G}({\bf x}_{\perp},{{\bf x}'}_{\perp})=
{f_{n-2}(\rho) \over \Omega_{n-1}}\, .
\ee
Here
\be \n{2.4a}
f_0(\rho)=\ln \rho~~,~
f_n(\rho)=-{1 \over n\rho^n}~~,~~n\geq1\, ,
\ee
$n=D-2$ is a number of the transverse directions, and $\Omega_{n-1}$
is the volume of the unit sphere $S^{n-1}$
\be \n{2.4d}
\Omega_{n-1}={2\pi^{n/2} \over \Gamma(n/2)}~~.
\ee

A solution to the equation (\ref{DEL}) is
\be
{h}_{\mu\nu}({\bf x}_{\perp})=-\kappa \int d{{\bf x}'}_{\perp}
{\cal G}_n(|{\bf x}_{\perp}-{{\bf x}'}_{\perp}|)t_{\mu\nu}({{\bf x}'}_{\perp})~~.
\ee
For a narrow beam of radiation the integration is performed over a
region of the size $\sim l$. If $R=|{\bf x}_{\perp}|\gg l$ one has
\be
|{\bf x}_{\perp}-{{\bf x}'}_{\perp}|\sim R-{ ({\bf x}_{\perp},{{\bf
x}'}_{\perp})\over R}\, .
\ee
By keeping the leading and sub-leading terms one obtains
\be \n{2.8}
{h}_{\mu\nu}({\bf x}_{\perp})=-{\kappa \over \Omega_{n-1}}
\left[f_{n-2}(R) \tau_{\mu\nu}-{f'_{n-2}(R) \over
R}x^b\sigma_{\mu\nu b}\right],
\ee
where
\be \n{2.9}
\tau_{\mu\nu}=\int d{\bf x}_{\perp}\, t_{\mu\nu}({\bf x})~~,
\ee
\be \n{2.10}
\sigma_{\mu\nu a}=\int d{\bf x}_{\perp}\, t_{\mu\nu}({\bf x}')~{x}^a~~.
\ee
Using relation (\ref{3.12a})
one finds that
\be \n{3.14a}
\tau_{uu}=E/L~~~.
\ee
Since $t_{ua}$ is a derivative of a function vanishing at infinity,
its integral $\tau_{ua}$ vanishes. The conservation of energy
implies the identity $\int d{\bf x}_{\perp}\,\partial^c t_{uc}x^bx^a=0$ and
the property $\sigma_{uab}=-\sigma_{uba}$.
Therefore,
\be \n{3.14b}
\sigma_{uab}=\sigma_{aub}=-\frac 12 J_{ab}/L~~~,
\ee
where we used the definition (\ref{3.13}).

We use the freedom of rotation of the transverse coordinates ${\bf x}_{\perp}^a$
in order to relate the coordinates to the bi-planes
determined by the antisymmetric tensor $J^{ab}$. We enumerate the
bi-planes by the index $i=1,\ldots,l$,  and assume that after a
proper rotation brings the antisymmetric matrix $J_{ab}$ to its canonical
form, the new coordinates $x_i=x^{2i+1}$ and $y_i=x^{2i+2}$  belong to the
$i$-th bi-plane. As earlier we denote by $J_i$ the corresponding
component of the angular momentum. In these coordinates the metric in
the weak field approximation takes the form
\[
ds^2=-du \, dv +\sum_{i=1}^l (dx_i^2+dy_i^2) +\epsilon dz^2
\]
\be \n{e1}
+2\Phi\, du^2 +2du
\sum_{i=1}^l A_i\, (y_i dx_i-x_i dy_i)\, ,
\ee
\be \n{3.16}
\Phi=-{\kappa E \over 2\Omega_{D-3}} \chi (u) f_{D-4}(R) ~~,
\ee
\be \n{3.17}
A_i=-{\kappa  \over 4\Omega_{D-3}}J_{i} \chi_{i}(u)
{1 \over R^{D-2}}~~.
\ee
In the even-dimensional spacetime $\epsilon=0$, and in the
odd-dimensional one it is equal to 1. According to the above definitions
\be
R^2=\sum_{i=1}^l (x_i^2+y_i^2) +\epsilon z^2\, .
\ee
It should be also emphasized that we
restored  the dependence on $u$ in the solution (\ref{e1}) by
introducing the profile functions $\chi(u)$ and $\chi_i(u)$ corresponding, respectively,
to $uu$ and $ux_i$, $uy_i$ components of the stress-energy tensor of the beam.
As we explained earlier, this can be done simultaneously in the source
and the field components. To preserve the meaning of $E$ and $J_i$ as
a total energy and total angular momentum components we must require
the following normalization conditions
\be\n{norm}
\int du \chi(u)= \int du \chi_i(u)=1\, .
\ee

It is convenient to rewrite the metric (\ref{e1}) in a different form.
Let us introduce polar coordinates $(r_i,\phi_i)$ in the $i$-th
bi-plane,
\be \n{3.20}
x_i=r_i\cos\phi_i~~,~~y_i=r_i\sin\phi_i\, .
\ee
Then the metric (\ref{e1}) takes the form
\be \n{lin_m}
ds^2_{\mbox{\tiny{even}}}= -du dv
+\sum_{i=1}^{l}( dr_i^2 +r_i^2 d\phi_i^2 ) +\epsilon dz^2
\ee
\[
 +2\Phi  du^2+{{\cal A} \over R^{D-2}}
\sum_{i=1}^l J_i \chi_i(u) r_i^2 d\phi_i   du \, ,
\]
where ${\cal A}= \kappa /(2\Omega_{D-3})$.

In the special case of the four-dimensional space-time the metric
(\ref{lin_m}) is
\[
ds^2=-du \, dv + dr^2+r^2d\phi^2
\]
\be \n{3.18}
-4 G\left[ 2E  \chi (u)\ln r ~du - J  \chi_1 (u)d\phi\right] du\, ,
\ee
In 4-dimensions $R=r$ and $J$
is the internal angular momentum (spin) of the beam.

\section{Gravitational field of a
relativistic gyraton: Exact higher dimensional solutions}

In this section we discuss the solution of the higher dimensional
Einstein equations describing the gravitational field of a
relativistic spinning beam-pulse of radiation (gyraton). In the
general case such a solution contains $l+1$ arbitrary functions of
the retarded time $u$, where $l$ is a number of independent bi-planes
of rotation. The form of the solution is slightly different for even
and odd number $D$ of the space-time dimensions. In this section we
assume that $D>4$. The case $D=4$, which requires special
consideration, will be considered in the next section.

In the even dimensional case, $D=2l+2$, the solution can be written
as follows \cite{zz}
\[
ds^2_{\mbox{\tiny{even}}}=- du dv +2\sum_{i=1}^l dz^i d\bar{z}^i
+(2\Phi+{\cal B}) du^2
\]
\be\n{4.1}
+ 2\sum_{i=1}^l (W_i
dz^i+\bar{W}_{i} d\bar{z}^i) du \, .
\ee
Here
\be\n{4.2}
W_i=-i p_i(u){\bar{z}^i\over R^{2l}}\hh
R^2=2\sum_{i=1}^l z^i \bar{z}^i\, ,
\ee
\be\n{4.3}
\Phi={\mu (u)\over R^{2(l-1)}}\hh
{\cal B}={1\over R^{4l-2}} \left[ \alpha_l {P^2\over R^{2}}+\beta_l
p^2\right]\, .
\ee
$z^i$, $\bar{z}^i$ are complex coordinates
and the bar denotes
complex conjugation.
The function $\mu(u)$ and $l$ functions $p_i(u)$ are arbitrary
functions of $u$. We also use the following notations
\be\n{4.5}
P^2=2\sum_{i=1}^l p_i^2(u) z^i\bar{z}^i
\hh
p^2=\sum_{i=1}^l p_i^2(u)\, ,
\ee
\be\n{4.6}
\alpha_l={l-2\over 2(l-1)}\hh
\beta_l={1\over 2(l-1)(2l-1)}\, .
\ee
In the appendix B we prove that this metric is Ricci-flat everywhere
outside $z^i=\bar{z}^i=0$, and hence it is a solution of
vacuum Einstein equations outside the source.

In order to explain the meaning of $\mu$ and $p_i$ it is
convenient to rewrite this metric in a form similar to
(\ref{lin_m}). For this purpose we denote
\be\n{4.7}
z^i={r_i\over \sqrt{2}} e^{i\phi_i}\, .
\ee
In these coordinates the metric (\ref{4.1}) takes the form
\be\n{4.8}
ds^2_{\mbox{\tiny{even}}}= -du dv
+\sum_{i=1}^{l}( dr_i^2 +r_i^2 d\phi_i^2 )
\ee
\[
 +(2\Phi + {\cal B})du^2+{2\over R^{D-2}}
\left[ \sum_{i=1}^l p_i(u) r_i^2 d\phi_i\right]  du \, .
\]
$\Phi$ and ${\cal B}$ are given by (\ref{4.3}) with
\be\n{4.9}
R^2=\sum_{i=1}^{l} r_i^2\hh
P^2=\sum_{i=1}^{l} p_i^2(u) r_i^2\, .
\ee

Let us suppose that the following integrals are finite
\be
m=\int_{-\infty}^{\infty} du \mu(u)\, ,
\ee
\be
j_i=\int_{-\infty}^{\infty} du p_i(u)\, ,
\ee
and denote
\be
\chi(u)=\mu(u)/m\hh
\chi_i(u)=p_i(u)/j_i\, .
\ee
By comparing the metric (\ref{4.8}) with a linearized solution
(\ref{lin_m}), one can conclude that $m$ is proportional to the total
energy $E$ of the pulse of radiation, while $j_i$ are proportional to
the independent angular momenta $J_i$ of the pulse
\be\n{en_an}
m= {\kappa E\over 2\Omega_{D-3}(D-4)}\hh
j_i={ \kappa J_i\over 4\Omega_{D-3}}
\ee
The functions $\chi(u)$ and $\chi_i(u)$ describe profiles of
distributions of the energy and angular momenta of the radiation
within the pulse.

In the odd dimensional case $D=2l+3$ the solution has the metric
\be\n{m_odd}
ds^2_{\mbox{\tiny{odd}}}=ds^2_{\mbox{\tiny{even}}}+dz^2\, ,
\ee
where $ds^2_{\mbox{\tiny{even}}}$ is given by (\ref{4.1}) and
\be\n{4.2a}
W_i=-i p_i{\bar{z}^i\over R^{(2l+1)}}\hh
R^2=2\sum_{i=1}^l z^i \bar{z}^i+z^2\, ,
\ee
\be\n{4.3a}
\Phi={\mu (u)\over R^{2l-1}}\hh
{\cal B}={1\over R^{4l}} \left[ \alpha_l {P^2\over R^{2}}+\beta_l
p^2\right]\, ,
\ee
\be\n{b.107}
\alpha_l={2l-3\over 2(2l-1)}\hh
\beta_l={1\over 2l(2l-1)}\, .
\ee
The function $\mu(u)$ and $l$ functions $p_i(u)$ are arbitrary
functions of $u$, and $P^2$ and $p^2$ are given by (\ref{4.5})
This metric in the radial coordinates (\ref{4.7}) has the form
(\ref{m_odd}) where $ds^2_{even}$ is now given by expression
(\ref{4.8}) with
\be
R^2=\sum_{i=1}^{l} r_i^2 +z^2\, .
\ee
As in the even dimensional case, the odd-dimensional solution has
$l+1$ arbitrary functions of $u$ which are related to the energy of
the pulse and its angular momenta by (\ref{en_an}).

\section{Gravitational field of a
relativistic gyraton: Exact solution in 4-dimensions}

As we already mentioned,  the case of 4-dimensional space-time is
special. in this section we derive a $4-D$ solution for the
gravitational field of a gyraton.
We use the following ansatz for the metric
\be \n{5.1}
ds^2=-du \, dv +d{\bf x}^2+(2\Phi\, du +2A_a\, dx^a) du\, ,
\ee
where $a,b=2,3$, ${\bf x}=(x^2,x^3)$, $d{\bf x}^2=(dx^2)^2+(dx^3)^2$,
$\Phi=\Phi({\bf x},u)$, and $A_a=A_a({\bf x},u)$. Straightforward but
rather long calculations give the following expressions for the
non--vanishing components of the Riemann curvature
\[
R_{u 2 u 2}=-\pa^2_2 \Phi +{1\over 2}  (\Rot {\bf
A})^2 +\pa_u \pa_2 A_2\, ,
\]
\be\n{5.2}
R_{u 3 u 3}=- \pa^2_3 \Phi +{1\over 2}(\Rot {\bf
A})^2 +\pa_u \pa_3 A_3\, ,
\ee
\[
R_{u 2 3 2}=-{1\over 2}\pa_2 \Rot {\bf A}\hh
R_{u 3 2 3}={1\over 2}\pa_3 \Rot {\bf A}\, ,
\]
\[
R_{u 2 u 3}=-\pa^2_{23}\Phi +{1\over 2} \pa_u (\pa_2
A_3 +\pa_3 A_2)\, .
\]
Here
\be\n{5.3}
\Rot {\bf A}=\pa_2 A_3-\pa_3 A_2\, ,
\ee
\be\n{5.4}
\Div {\bf A}=\pa_2 A_2 +\pa_3 A_3\, ,
\ee
\be
\lap \Phi =(\pa^2_2 +\pa^2_3)\Phi\, .
\ee

Calculations also give the following non-vanishing
components of the Ricci tensor
\[
R_{uu}=- \lap \Phi +{1\over 2}(\Rot {\bf
A})^2 +\pa_u \Div {\bf A}\, ,
\]
\be\n{5.5}
R_{ua}={1\over 2}\epsilon_{ab}\pa_b \Rot {\bf A}\, ,
\ee
where $\epsilon_{ab}$ is a $2-D$ anti-symmetric tensor, $\epsilon_{23}=-\epsilon_{32}=1$.

We suppose that everywhere outside the source at $x^2=x^3=0$ the
space-time is vacuum. Thus, outside the source one has
\be\n{5.6}
\lap \Phi=0\hh
\Div {\bf A}=0\hh
\Rot {\bf A}=0\, .
\ee
Denote $r=\sqrt{(x^2)^2+(x^3)^2}$.
Solutions of the equations (\ref{5.6}) which are regular at $r\to
\infty$ are
\be\n{5.7}
\Phi=-\mu(u)\ln r\hh
A_a={p(u)\over r^2}\epsilon_{ab} x^b\, .
\ee

Strictly speaking, $\Phi$ and ${\bf A}$ are distributions which obey
the following inhomogeneous equations
\be\n{fdel}
\lap \Phi=-2\pi \mu(u) \delta ({\bf x})\, ,
\ee
\be\n{adel}
\Rot {\bf A}=-2\pi p(u) \delta ({\bf x})\, .
\ee
It is interesting to note that the equations for
$\Phi$ and ${\bf A}$ coincide with the equations of 2-dimensional
electrodymanics. In this analogy $\Phi$ is similar to the Coulomb
potential of a point-like charge, while ${\bf A}$ is similar to the
vector potential for the axisymmetric magnetic field. An unusual
property of these fields is that the value of the charge and the flux
of the magnetic field depend on the "time" $u$.
In more general terms, this interpretation reflects a
well-known electromagnetic analogy of the gravitational interaction
(see e.g. \cite{Mash}).

Let us introduce the definitions
\be\n{mp}
\mu(u)=m\chi(u)~~,~~p(u)=-j {\chi}_1(u)~~,
\ee
where $\chi(u)$ and ${\chi}_1(u)$ are profile functions  for the
energy and angular momentum distributions, respectively, obeying the
normalization conditions (\ref{norm}). We also put
$$
m=4GE~~,~~j=2GJ~~~.
$$
Then in the polar coordinates $x^2=r\cos\phi$, $x^3=r\sin\phi$ the
metric (\ref{5.1}) takes the form
\[
ds^2=-du \, dv +r^2d\phi^2 +dr^2
\]
\be \n{5.8}
 -4G(2E \chi(u) \ln r ~du -J{\chi}_1(u) d\phi) du\, .
\ee
By comparison with the weak field approximation (\ref{3.18}) one can
conclude that $E$ and $J$ are the energy and the angular momentum of
the beam, respectively. The metric (\ref{5.8}) belongs to the Kundt's
class of metrics (see e.g. \cite{ES}).

The 4-dimensional metric $g_{\mu\nu}$ is
linear in $E$ and $J$. However
$\sqrt{-g}$ and $g^{\mu\nu}$ contain terms proportional to $J^2$.

The form of the metric (\ref{5.8}) is invariant under the following
transformations
\[
u^*= L u+b\, ,\hspace{0.2cm}
v^*= L^{-1}v\, ,
\]
\be\n{5.8a}
\chi^*(u^*)= L^{-1}\chi((u^*-b)/L)\, ,
\ee
\be
\tilde{\chi}^*(u^*)= L^{-1}\tilde{\chi}((u^*-b)/L)\, ,
\ee
where $L$ and $b$ are constants.

Equation $\Rot {\bf A}=0$ implies that at least locally
\be\n{5.9}
{\bf A}=\nabla \Theta\, .
\ee
Let us denote
\be\n{5.10}
\Sigma= \Phi-\pa_u \Theta\, .
\ee
Then for the Ricci flat metric the non-vanishing components of the
curvature can be written in the following compact form
\be\n{5.11}
R_{u a u b}=-\pa^2_{ab}\Sigma \,.
\ee

Let us emphasize that the obtained vacuum solution is valid only
outside the beam-pulse. On the beam axis it is singular.  One can
formally define  an effective stress-energy tensor  for the  solution
(\ref{5.8}) as
\be
T_{\mu\nu}=(8\pi G)^{-1}(R_{\mu\nu}-\frac 12 g_{\mu\nu}R) \, .
\ee
It can be shown that the Ricci scalar is identically zero. Thus, the
only non-trivial components of the stress tensor are
\be\n{5.12}
T_{uu}=E\chi(u)\delta({\bf x})+\pi G J^2 \chi_1^2(u)\delta^2({\bf x})~~.
\ee
\be\n{5.13}
T_{ua}= \frac J4\chi_1(u)\epsilon_{ab}\partial_b \delta({\bf x})~~.
\ee
To obtain these relations we used (\ref{5.5}) and (\ref{5.7}) and
took (\ref{adel}) into account. The term $\delta^2({\bf x})$  on the
right hand side of (\ref{5.12}) indicates that in the presence of
spin one must consider spatially distributed sources. In the weak
field approximation the second term on the right hand side of
(\ref{5.12}) should be omitted. Then $T_{uu}$, $T_{ua}$ take the form
of components of the stress-energy tensor of an infinitely narrow
beam with the energy $E$ and the
angular momentum tensor $J_{ab}=\frac 12 \epsilon_{ab} J$.

\section{Particles and light motion in the field of gyraton}

In this section we describe briefly the gravitational force of a
relativistic gyraton acting on particles and light rays. We restrict
ourselves by considering only the case of 4 dimensions.

Non-vanishing Christoffel symbols for the metric (\ref{5.8}) are
\be
\Gamma_{uu}^{\,\,v}=2{\mu}' \ln(r)+{2p{p}'\over r^2}\hh
\Gamma_{uu}^{\,\,r}={\mu\over r}\hh
\Gamma_{uu}^{\,\,\phi}={{p}'\over r^2}\, ,
\ee
\[
\Gamma_{ur}^{\,\,v}={2\mu\over r}\hh
\Gamma_{r\phi}^{\,\,v}={2p\over r}\hh
\Gamma_{r\phi}^{\,\,\phi}={1\over r}\hh
\Gamma_{\phi\phi}^{\,\,r}=- r\, ,
\]
where $\mu(u)$ and $p(u)$ are given by (\ref{mp}).
In this section we shall use a notation ${a}'=\pa_u
a$.

Since $\Gamma_{\mu\nu}^{\,\,u}=0$ one can choose $u$ as an affine
parameter. We shall use this choice. The equations of motion in the
2-dimensional plane orthogonal to the direction of the motion are
\be
{r}''+{\mu\over r}-r{\phi '}^2=0\, ,
\ee
\be
{\phi}''+{{p}'\over r^2}+{2{\phi}'{r}'\over r}=0\, .
\ee
Because of the axial symmetry of the metric the
second equation allows an integral of motion which we denote by $p_0$
\be\n{phi}
{\phi}'=-{(p-p_0)\over r^2}\, .
\ee
This integral of motion is a quantity connected with the conserved
angular momentum. The radial equation can be written as follows
\be\n{rad}
r''+{\mu\over r}-{(p-p_0)^2\over r^3}=0\, .
\ee
Instead of the last equation of motion for $v(u)$ it is more
convenient to use its integral which follows from the normalization
condition
\be
g_{\mu\nu}{x^{\mu}}'{x^{\nu}}'=\epsilon\, ,
\ee
where $\epsilon=0$ for light rays and $\epsilon=-1$ for particles.
This equation gives
\be
-v'+r^2{\phi'}^2+{r'}^2-2\mu \ln r +2p\phi'=\epsilon\, .
\ee
Using (\ref{phi}) it can be rewritten  as
\be
-v'-{p^2\over r^2}+{r'}^2-2\mu \ln r =\epsilon\, .
\ee

For $\mu=p=0$ the equations of motion can be easily integrated. Using
an ambiguity in the integration constant the trajectory for this case
is
\be
r={p_0\over \cos\phi}\, ,
\ee
that is, it is a straight line passing at the distance from the center
(impact parameter) equal to $p_0$.

\begin{figure}[h]
\begin{center}
\includegraphics[height=3.0cm,width=5cm]{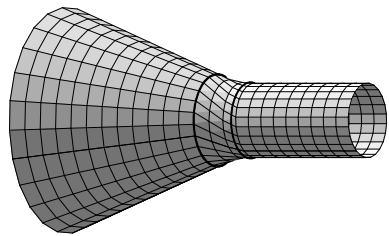}
\caption{The relativistic gyraton with $m=0$, $j=0.7$ passes through the center of a set
of particles initially located at the radius 1. The gravitational
field of the gyraton is turned on during the interval $(-1/2,1/2)$
(two solid circular curves). During this interval the particles
position is twisted. After this the particles are moving
radially outward with a uniform speed.}
\label{f1}
\end{center}
\end{figure}

\begin{figure}[h]
\begin{center}
\includegraphics[height=1.0cm,width=5cm]{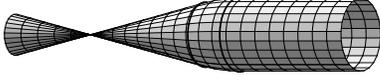}
\caption{Motion of particles for the gyraton parameters $m=0.25$,
$j=0$. For this case there is no twist, and particles move to the center
with constant radial velocity. They pass the center, $r=0$ after the
gyraton, and passing the caustic starts their expansion.}
\label{f2}
\end{center}
\end{figure}

\begin{figure}[h]
\begin{center}
\includegraphics[height=1cm,width=5cm]{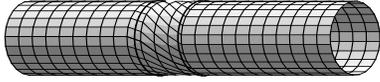}
\caption{A special case ($m=1$ $j=1$) when the centrifugal repulsion
compensates exactly the "Newtonian" attraction.}
\label{f3}
\end{center}
\end{figure}

\begin{figure}[h]
\begin{center}
\includegraphics[height=1cm,width=5cm]{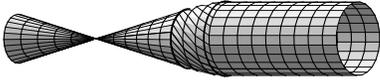}
\caption{The case when both of the parameters $m$ and $j$ do not
vanish, but the attraction dominates ($m=1.5$, $j=1$).}
\label{f5}
\end{center}
\end{figure}

\begin{figure}[h!]
\begin{center}
\includegraphics[height=2cm,width=5cm]{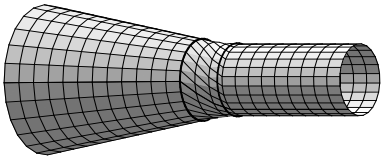}
\caption{The case when both of the parameters $m$ and $j$ do not
vanish, but the repulsion dominates ($m=0.7$, $j=1$).}
\label{f4}
\end{center}
\end{figure}

\begin{figure}[h]
\begin{center}
\includegraphics[height=3.5cm,width=5cm]{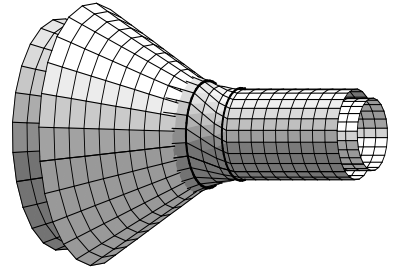}
\caption{A special case when the repulsion dominates ($m=0$, $j=0.6$).
After the gyraton passes throughout the center, the particles
located at the inner circle  $r_1=0.8$ get higher positive radial
velocity than the particle which initially were located at  $r_0=1.0$.
The inner particles (their worldsheet is shown by a light surface)
cross the outer particles (darker surface). At the moment of the
intersection a caustic is formed.}
\label{f6}
\end{center}
\end{figure}

In order to illustrate the action of the relativistic gyraton on the
motion of particles and light rays, let us consider a special case
when initially their equation of motion in the plane perpendicular to
the direction of motion  was
\be
\phi=\phi_0=0\hh r=r_0=\mbox{const}\, .
\ee
For
this case $p_0=0$. Using equation (\ref{phi}) one can see that when a
gyraton passes near the particle it effectively imparts an angular
momentum to it which is proportional to $p(u)$. The radial equations show
that the "mass" term $\mu(u)$ produces an acceleration directed
towards the
center, while the angular momentum term effectively produces an
acceleration directed away from the center, which is similar to the usual
centrifugal force. Consider a set of particles located on a circle
of initial radius $r_0$ orthogonal to the direction of motion of
the gyraton. As a result of action of a gyraton moving though the
center of the circle, the trajectories of the particles are twisted
and later either converge to the center or expand. The character of
the motion depends on which term in the equation (\ref{rad}) dominates.
The pictures~\ref{f1}--\ref{f6} illustrate this.

We consider a special case when both $\mu$ and $p$ have the same
step-like form. Using an ambiguity (\ref{5.8a}) in the choice of the
coordinate $u$ we can write
\be
\mu(u)=m \chi(u)\hh
p(u)=-j \chi(u)~~~,
\ee
\be\n{step}
\chi(u)=\vartheta(1/2-u)+\vartheta(1/2+u)-1\, .
\ee
The step-function $\chi$ vanishes before $u=-1/2$ and after $u=1/2$,
and is equal to 1 between these values. It is normalized so that
\be
\int_{-1/2}^{1/2} \chi(u) du=\int_{-1/2}^{1/2} \chi^2(u) du=1\, .
\ee

For this choice the gravitational field of the gyraton affects the
particles during the interval $(-1/2,1/2)$. Figure~\ref{f1} shows the
motion of the particles forming the circle for $m=0$ and $j=1$.  The
initial radius of the circle is chosen to be 1. The coordinate $u$
grows from the right to the left. In this and next figures, two solid
circular lines on the surface correspond to the moments $u$ when the
gravitational field of the gyraton is  switched on ($u_0=-1/2$, the
right circle) and the moment when it is switched off ($u_1=1/2$, the
left one). Between these two moments the particles trajectories are
twisted. After switching off the gravitational field the particles
are moving radially with some positive constant radial velocity.

Figure~\ref{f2} shows the motion of the particles for the other special case
$m=0.25$, $j=0$. As the result of the attraction to the moving
gyraton the radius of the circle (which was originally equal to 1)
shrinks from its original value. After $u=1/2$ the particles have a
constant negative radial velocity. They pass the point $r=0$ some time
after the gyraton was there. After a conical caustic
at this point the circle starts its linear expansion.

Figure~\ref{f3} illustrates a case when the "Newtonian" attraction is
exactly compensated by the "centrifugal" repulsion generated by the
rotation of the gyraton. After passing the gyraton the particles
remain at the same radius $r=r_0=1$.

In the general case either "Newtonian" attraction or "centrifugal"
repulsion dominates. Figures~\ref{f4} and \ref{f5} illustrate these
two options.

The "centrifugal force" term is proportional to $r^{-3}$. It means
that in the absence of the "Newtonian" attraction, $\mu=0$, the
outward acceleration of particles at smaller radius is bigger than
the acceleration at bigger radius. Consider the evolution in time of
two circles, one with the original radius $r_0$ and the other with
$r_1<r_0$. Since the final positive radial velocity  is bigger in the
second case, a surface $\Sigma_1$ representing the second circle
motion (light surface at Figure~\ref{f6}) crosses from the inside the
(darker) surface $\Sigma_0$ at some moment of time $u$.

In a general case such points of intersection form a caustic curve
$u=U(r)$. If the interval during which the field of the gyraton is
switched on is much smaller than the time formation of the caustics
one can define the function $F$ as follows. Denote by
\be\n{jj}
\tilde{j}^2=j^2 \int_{-\infty}^{\infty} du \, \chi^2(u)\, .
\ee
For the step function
$\chi$ given by (\ref{step}) $\tilde{j}=j$. By integrating the radial equation
(\ref{rad}) over a (short) interval when the interaction is switched on
and assuming that during this interval the radius remains
approximately constant one obtains the following relation
\be\n{Dr}
\Delta[r']={\tilde{j}^2\over r_0^3}-{m\over r_0}\, .
\ee
Here $\Delta[r']$ is the change of the radial velocity.
In the same approximation
\be
\Delta[\phi]={j^2\over r_0^2}\, .
\ee

We assume that
particles before the gyraton passing nearby were at a fixed radius.
After the gyraton has passed, the equation of the particle motion is
\be\n{rel}
r=F(r_0,u)=r_0+\left({\tilde{j}^2\over r_0^3}-{m\over r_0}\right) u\, .
\ee
The condition for caustic formation is $\pa_{r_0}F=0$. By solving
this equation one obtains the following condition for the caustic line
formation
\be\n{uu}
u=U(r_0)={r_0^4\over 2\tilde{j}^2-mr_0^2}\, .
\ee
Substituting this relation into (\ref{rel}) one obtains
\be\n{rr}
r={r_0(3\tilde{j}^2-2mr_0^2)\over 2\tilde{j}^2-mr_0^2}\, .
\ee
The equations (\ref{uu}) and (\ref{rr}) describe the caustic equation
in the parametric form. Since $r$ must be positive, $r_0^2\le
3\tilde{j}^2/(2m)$. When $r_0$ is close to the limiting value
$\sqrt{3}\tilde{j}/\sqrt{2 m}$ $u$ becomes
\be\n{uuu}
u\approx {9\tilde{j}^2\over 2m^2}\, .
\ee

We remind the reader that the above estimations are given in a
special frame where the duration of the pulse is 1. Using the scaling
law (\ref{5.8a}) it is possible to rewrite the results in an
arbitrary frame where the duration of the pulse is $L$. It is
sufficient to notice that under this scaling $m$ and $r$ do not
change, while $\tilde{j}\to \tilde{j}/\sqrt{L}$. In particular, under
this transformation the relation (\ref{uu}) remains invariant.

Equation (\ref{Dr}) allows one to make the following general
conclusion. "Centrifugal" repulsion compensates the "Newtonian"
attraction when
\be\n{r00}
r_0\sim {j\over \sqrt{mL}}\sim {\sqrt{G} J\over \sqrt{EL}}\,
.
\ee

It should be emphasized that a realistic beam pulse of spinning
radiation has a finite cross section, so that the formulas and
approximations used above should be additionally tested for their
consistency.

\section{Summary and Discussions}

Let us summarize the obtained results. The beam-pulses of spinning
radiation besides energy have also internal angular momentum. As a
result, the metric for the gravitational field of such
relativistic gyratons in addition to the "Newtonian" part contains
non-diagonal terms responsible for the gravitomagnetic effects. In
the weak field approximation the gravitational field of the
relativistic gyraton is related to the boosted Lense-Thirring metric
in the limit when the boost parameter is infinitely large. Since the
angular momentum in this limit remains finite, the weak field boosted
solution is not sufficient to generate an exact solution.
In this paper we obtained the exact
solutions for the relativistic gyraton in an arbitrary number $D$ of
spacetime dimensions. These solutions
contain $[D/2]$ arbitrary profile functions of one parameter $u$. They
describe the energy density, $\mu (u)$, and angular momentum density,
$p_i(u)$, along the beam.

When a relativistic gyraton passes near a particle, the motion of
the latter is affected by the "Newtonian" attraction generated by the
energy $\mu$ and by an induced `centrifugal' force, generated by the
angular momentum of the gyraton. We explicitly demonstrated this
for the four-dimensional case but this conclusion is of general
nature.

The condition (\ref{r00}) for the radius where repulsive and
attractive forces are equal is obtained for classical sources. It is
interesting to apply it (at least formally) to the case of a single
quantum. For the quantum with wavelength $\lambda$ and spin $s$ one
has $E=\hbar/\lambda$ and $J=s \hbar$ and the condition (\ref{r00})
takes the form
\be r_0\sim s l_{Pl} \sqrt{ {\lambda\over L}}\, .
\ee
Here $l_{Pl}=\sqrt{\hbar G}$ is the Planck length. At the
threshold of the mini-black-hole production when both the
center-of-mass energy of particles and their impact parameter have
the Planckian scale,  this relation shows that the spin-orbit
interaction becomes comparable with the "Newtonian" attraction for a
quantum with $L\sim \lambda$.

In case when both particles have a
spin, besides the spin-orbit interaction there exists  also an
additional spin-spin interaction. One can expect that both effects
might be important in the Planckian regime and change the estimated
cross-section of mini black hole formation. In order to study this
problem one needs to find the metric for ultra-relativistic particles
with a spin. The solutions obtained in this paper for the
relativistic gyraton can be a good starting point in this investigation.

\noindent
\section*{Acknowledgment}\noindent
This work was supported by the Natural Sciences and Engineering
Research Council of Canada and by the Killam Trust.
The authors also kindly acknowledge the support from  the NATO
Collaborative Linkage Grant (979723). The authors are grateful to
Werner Israel, Bahram Mashhoon and Andrei Zelnikov for stimulating
discussions and comments.

\medskip

\appendix

\section{Boost of a compact source}

In the case when $\chi(u)=\chi_i(u)=\delta(u)$ metric (\ref{e1})--(\ref{3.17}) can be obtained
from the metric of a source by the boost.
Consider a compact source in a flat space-time which is at rest in the frame
of reference with coordinates $\bar{x}^\mu$, such that $\bar{x}^1=\bar{t}$
($\bar{t}$ is time coordinate)
$\bar{x}^2=\bar{\xi}$, $\bar{x}^a=x^a$, $a=3,..,D$.
We call $x^a$ transverse coordinates and denote them by the vector ${\bf x}_{\perp}$.
The source has the mass
$$
M=\int d\bar{\xi} d{\bf x}_{\perp}T_{00}~~,
$$
and the angular momentum tensor
$$
J_{\mu\nu}=\int d\bar{\xi} d{\bf x}_{\perp}(x^\mu T^\nu_0-x^\nu T^\mu_0)~~,
$$
where $T_{\mu\nu}$ is the stress-energy tensor of the source.
It is assumed that the source rotates only in transverse directions, hence the only non-trivial
components of the angular momentum tensor are $J_{ab}$.
The number of rotation bi-planes is $l=[(D-2)/2]$.
As before we work in the
transverse coordinates corresponding to the rotation bi-planes, where $J_{ab}$ has a
block-diagonal form, and denote $J_i$  the momentum
in the $i$-th bi-plane.

In the weak field approximation the gravitational field of the source
is given by the metric \cite{MyPe}
$$
ds^2=-d\bar{t}^2+d\bar{\xi}^2+d{\bf x}_{\perp}^2+
$$
$$
2\tilde{\Phi}\,\left(d\bar{t}^2+{1 \over D-3}(d\bar{\xi}^2+d{\bf x}_{\perp}^2)\right)
+
$$
\be \n{a.1}
2d\bar{t}\sum_{i=1}^l\tilde{A}_i\, (y_i dx_i-x_i dy_i) \, ,
\ee
where
$$
d{\bf x}_{\perp}^2=\sum_{i=1}^l (dx_i^2+dy_i^2) +\epsilon dz^2~~,
$$
\be \n{a.2}
\tilde{\Phi}(\rho)=-{\kappa M (D-3)\over 2(D-2)\Omega_{D-2}} f_{D-3}(\rho) ~~,
\ee
\be \n{a.3}
\tilde{A}_i(\rho)=-{\kappa  \over 4\Omega_{D-2}}{J_i \over \rho^{D-1}}~~.
\ee
Functions $f_{D-3}$ are defined by (\ref{2.4a}),
$$
\rho^2=\bar{\xi}^2+{\bf x}_{\perp}^2=\bar{\xi}^2+R^2~~~,
$$
and $\epsilon=0$ if the number of transverse directions is even.

Let us go to a frame of reference which moves
backwards along $\xi$-axis with the velocity $\beta$ (we work in the system
of units where the velocity of light $c=1$).
Let $\xi$, $t$, $u=t-\xi$, $v=t+\xi$ be the corresponding coordinates
in the moving frame,
\be \n{a.4}
\bar{\xi}=\gamma(\xi-\beta t)~~,~~\bar{t}=\gamma(t-\beta\xi)~~,
\ee
where $\gamma=(1-\beta^2)^{-1/2}$. By using (\ref{a.4})
one can consider the limit $\beta\rightarrow 1$. In this limit
the source is moving with respect to the new frame of reference with
the velocity of light.
Let $\triangle_{\perp}$  be the Laplace operator in the transverse directions and
\be \n{a.7}
\triangle=\triangle_{\perp}+\partial_{\bar{\xi}}^2~~.
\ee
In the limit of infinite boost operators
$\triangle$ and $\triangle_{\perp}$ coincide up to terms $O(\gamma^{-2})$.
According to definitions of Green's functions (\ref{G1}), (\ref{G2}),
\be \n{a.8}
\triangle f_{D-3}(\rho)=
\Omega_{D-2}\delta(\bar{\xi})\delta({\bf x}_{\perp})~~,
\ee
\be \n{a.9}
\triangle_{\perp} f_{D-4}(R)=\Omega_{D-3}\delta({\bf x}_{\perp})~~.
\ee
In the limit of infinite boost  the following relations hold
\be \n{a.5}
\lim_{\beta\rightarrow 1}(\gamma f_1(\rho))=f_1(u)+2\delta(u)f_0(R)~~,
\ee
\be \n{a.6}
\lim_{\beta\rightarrow 1}(\gamma f_{D-3}(\rho))=\delta(u){\Omega_{D-2} \over
\Omega_{D-3}}f_{D-4}(R)~~,~~D>4~~,
\ee
$$
\lim_{\beta\rightarrow 1}\left(\gamma {x_i \over \rho^{D-1}}\right)=
$$
\be \n{a.10}
\lim_{\beta\rightarrow 1}\left(\gamma \partial_{x_i} f_{D-3}(\rho)\right)=
\delta(u){\Omega_{D-2} \over
\Omega_{D-3}}{x_i \over R^{D-2}}~~,
\ee
which can be found with the help of (\ref{a.8}), (\ref{a.9})
if one has takes into account that
$\lim_{\beta\rightarrow 1}(\gamma \delta(\bar{\xi}))=\delta(u)$.

By using (\ref{a.5})--(\ref{a.10}) one finds that
in the limit of the infinite boost metric (\ref{a.1}) takes the form
(\ref{e1})--(\ref{3.17}) provided that the mass of the source behaves in this limit as
$$
M={E \over \gamma}
$$
while components of the angular momentum $J_i$ remain finite.
To get the gyraton metric (\ref{5.8})
from the boost in four dimensions one has to make an
additional coordinate transformation $\tilde{v}=v-4GE\ln u/u_0$,
where $u_0$ is a constant.

\section{Metric ansatz and curvature calculations}

\subsection{General formulas}

We define the metric to be
\be\n{b.1}
ds^2=\eta_{\h{\mu}\h{\nu}}\w{\mu} \w{\nu}\, .
\ee
Here $\w{\mu}=\w{\mu}_{\nu} dx^{\nu}$ are basic forms and
$\eta_{\h{\mu}\h{\nu}}$ is a non-degenerate matrix with constant
coefficients. We denote $e_{\h{\mu}}=e_{\h{\mu}}^{\nu}\pa_{\nu}$
basic vectors dual to the basic forms
\be\n{b.2}
e_{\h{\mu}}^{\nu} \w{\lambda}_{\nu}=\delta^{\h{\mu}}_{\h{\lambda}}\, .
\ee

Let us define
\be\n{b.3}
\lambda_{\h{\mu}\h{\nu}\h{\lambda}}=e_{\h{\mu}}^{\epsilon}
e_{\h{\lambda}}^{\rho} [\omega_{\h{\nu}
\epsilon,\rho}-\omega_{\h{\nu}\rho,\epsilon}]\, .
\ee
These coefficients possess the following property
\be\n{b.4}
\lambda_{\h{\mu}\h{\nu}\h{\lambda}}=-\lambda_{\h{\lambda}\h{\nu}\h{\mu}}\,
.
\ee

The Ricci rotation coefficients are defined as
\be\n{b.5}
\gamma_{\h{\mu}\h{\nu}\h{\lambda}}={1\over 2}
[\lambda_{\h{\mu}\h{\nu}\h{\lambda}}+
\lambda_{\h{\lambda}\h{\mu}\h{\nu}}
-\lambda_{\h{\nu}\h{\lambda}\h{\mu}}]\,
.
\ee
The Riemann tensor can be written by using the Ricci rotation
coefficients as follows (see e.g. \cite{Chand})
\[
R_{\h{\mu}\h{\nu}\h{\lambda}\h{\rho}}=
-\gamma_{\h{\mu}\h{\nu}\h{\lambda},\h{\rho}}+
\gamma_{\h{\mu}\h{\nu}\h{\rho},\h{\lambda}}
\]
\be\n{b.6}
+
\gamma_{\h{\nu}\h{\mu}\h{\epsilon}}
[ \gamma_{\h{\lambda}\  \h{\rho}}^{\ \ \h{\epsilon}}-
 \gamma_{\h{\rho}\  \h{\lambda}}^{\ \ \h{\epsilon}}]
\ee
\[
+
\gamma_{\h{\epsilon}\h{\mu}\h{\lambda}}
\gamma_{\h{\nu}\ \h{\rho}}^{\ \ \h{\epsilon}}-
\gamma_{\h{\epsilon}\h{\mu}\h{\rho}}
\gamma_{\h{\nu}\ \h{\lambda}}^{\ \ \h{\epsilon}}\, .
\]
Here $(\ldots)_{,\h{\mu}}=e^{\nu}_{\h{\mu}}(\ldots)_{,\nu}$. Finally the
Ricci tensor is
\be\n{b.7}
R_{\h{\mu}\h{\lambda}}=\eta^{\h{\nu}\h{\rho}}
R_{\h{\mu}\h{\nu}\h{\lambda}\h{\rho}}\, .
\ee

\subsection{Even-dimensional space-time}

Formulas for even and odd dimensional cases are slightly different.
Consider first the even dimensional case and denote $l=(D-2)/2$. We
shall use two real coordinates, $x^1=u$ and $x^2=v$. The other
coordinates $x^a$, $a=3,\ldots,D$ are complex. We shall use the
following notations for this set of complex
coordinate $x^i=\zeta^i$, $x^{\bar{i}}=\bar{\zeta}^{i}$, where
$i=3,5, \ldots, 2l+1$ and $\bar{i}=i+1$. In this notations a partial
derivative $A_{,a}$ denotes the following set of
partial derivatives
\be\n{b.8}
A_{,i}=\pa_{\zeta^i} A\hh A_{,\bar{i}}=\pa_{\bar{\zeta}^i} A\, .
\ee
We shall use the same convention for the indices connected with the
basic vectors and forms.
We denote
\be\n{b.9}
\eta^{ab}=\sum_{i}
[\delta^a_{i}\delta^b_{\bar{i}}+\delta^a_{\bar{i}}\delta^b_{i}]\, .
\ee
Using this notation we can write the metric in the form
\be\n{b.10}
ds^2=-2\omega^{\h{1}}\omega^{\h{2}}+2\sum_i
\w{i}\bw{i}\, .
\ee

We adopt the following ansatz for the metric.
Let us denote
\be\n{b.11}
\omega^{\h{1}}=du\hh
\omega^{\h{2}}=\frac 12 dv-Q du\, ,
\ee

\be\n{b.12}
\omega^{\h{i}}=d\zeta^i+{W}^i du\hh
\omega^{\h{\bar{i}}}\equiv \bar{\omega}^{\h{i}}=d\bar{\zeta}^i+\bar{W}^i du\, .
\ee
Here $Q=Q(u,\zeta_i,\bar{\zeta}_i)$ and
$W^i=W^i(u,\zeta_i,\bar{\zeta}_i)$.
We also define
\be\n{b.13}
W_i={W}^{\bar{i}}=\bar{W}^i\hh
W_{\bar{i}}=\bar{W}_i={W}^{i}\, .
\ee

The differentials of the coordinates can be expressed in terms of
basic forms as follows
\be\n{b.14}
du=\w{1}\hh
dv=2(\w{2}+Q\w{1})\, ,
\ee
\be\n{b.15}
d\zeta^i=\w{i}-{W}^i \w{1}\hh
d\bar{\zeta}^i=\bw{i}-\bar{W}^i \bw{1}\, .
\ee

The  corresponding vector basis is
\be\n{b.16}
e_{\h{1}}=\pa_u-{\cal D}+2 Q \pa_v\hh
e_{\h{2}}=2 \pa_v\, ,
\ee
\be\n{b.17}
e_{\h{i}}=\pa_{\z^i}\hh
e_{\h{\bar{i}}}\equiv \bar{e}_{\h{i}}=\pa_{\bz^i}\, ,
\ee
\be\n{b.18}
{\cal D}=\sum_i ({W}^{i} \pa_{\z^i}+\bar{W}^{i} \pa_{\bz^i})\, .
\ee

The metric is
\[
ds^2=- du dv +2\sum_i d\zeta^i d\bar{\zeta}^i+2(Q+\sum_i W_i W_{\bar{i}})
du^2\, .
\]
\be\n{b.19}
+ 2\sum_i (W_i
d\zeta^i+W_{\bar{i}} d\bar{\zeta}^i) du \, .
\ee

Calculations give the following non-vanishing components of
$\lambda_{\h{\mu}\h{\nu}\h{\lambda}}$
\be\n{b.20}
\lambda_{\h{1}\h{1}\h{a}}=Q_{,a}\, .
\ee
\be\n{b.21}
\lambda_{\h{1}\h{a}\h{b}}=W_{a,b}\, .
\ee
Non-vanishing components of the rotation coefficients are
\be\n{b.22}
\gamma_{\h{1}\h{a}\h{1}}= Q_{,a}\, ,
\ee
\be\n{b.23}
\gamma_{\h{1}\h{a}\h{b}}= {1\over 2}(W_{a,b}+W_{b,a})\, ,
\ee
\be\n{b.24}
\gamma_{\h{a}\h{b}\h{1}}= {1\over 2}(W_{a,b}-W_{b,a})\, ,
\ee

Notice that the rotation coefficients do not vanish only if their
first or the third index is equal to 1. If any of its indices is 2 the
rotation coefficient vanishes. Using these properties and the
definition of the Ricci tensor (\ref{b.7}) it is possible
to show that
\be\n{b.25}
R_{\h{a}\h{b}}=R_{\h{2}\h{a}}=R_{\h{a}\h{2}}=0\, .
\ee
Non-vanishing coefficients of the Ricci tensor are
\be\n{b.26}
R_{\h{1}\h{1}}=-\lap Q+ Z_{,u}-{\cal D} Z-U\, ,
\ee
\be\n{b.27}
R_{\h{1}{a}}=-{1\over 2} \lap W_{a}+{1\over 2}Z_{,a}\, .
\ee
Here $\lap=\eta^{ab}\pa_a\pa_b$ and
\be\n{b.28}
Z=\eta^{ab}W_{a,b}=\sum_i (W_{i,\bar{i}}+W_{\bar{i},i})\, ,
\ee
\be\n{b.29}
U={1\over 2}(\eta^{ac}\eta^{bd}+\eta^{ad}\eta^{bc}) W_{a,b}W_{c,d}\, .
\ee

To obtain a solution we choose
\be\n{b.30}
W_i=-i p_i{\bar{\zeta}^i\over R^{2l}}\, ,
\ee
\be\n{b.31}
R^2=2\sum_i \zeta^i \bar{\zeta}^i\, .
\ee
Here  $p_i=p_i(u)$ are arbitrary functions of $u$.
We demonstrate now that for this choice the components
$R_{\h{1}\h{a}}$  of Ricci tensor vanish.

It is easy to check that
\be\n{b.32}
W_{i,j}=2il p_i {\bar{\zeta}^i \bar{\zeta}^j\over R^{2(l+1)}}\, ,
\ee
\be\n{b.33}
W_{\bar{i},\bar{j}}=-2il p_i {{\zeta}^i {\zeta}^j\over R^{2(l+1)}}\, ,
\ee
\be\n{b.34}
W_{i,\bar{j}}={i \over R^{2l}}\left[ 2l p_i{\bar{\zeta}^i
{\zeta}^j\over R^{2}}-p_i \delta_{ij}
\right]\, ,
\ee
\be\n{b.35}
W_{\bar{i},{j}}=-{i \over R^{2l}}\left[ 2l p_i{{\zeta}^i
\bar{\zeta}^j\over R^{2}}-p_i \delta_{ij}
\right]\, .
\ee

It is easy to see that for each $i$, $W_{i,\bar{i}}+W_{\bar{i},i}=0$,
and hence $Z=0$.
We also have (outside a singular point $R=0$)
\be\n{lap1}
\lap \left({1\over R^{2m}}\right)={4m(m-l+1)\over R^{2(m+1)}}\, ,
\ee
\be\n{lap2}
\lap \left({\zeta^i\over R^{2m}}\right)={4m(m-l)\zeta^i\over R^{2(m+1)}}\, ,
\ee
\be\n{lap3}
\lap \left({\zeta^i\bar{\zeta}^i\over R^{2m}}\right)={2\over R^{2m}}\left[
1+{2m(m-l-1)\zeta^i\bar{\zeta}^i\over R^2} \right]\, .
\ee
The equation (\ref{lap2}) implies
\be\n{b.39}
\lap W_i=\lap W_{\bar{i}}=0\, .
\ee
Thus $R_{\h{1}\h{a}}=0$ and $R_{\h{1}\h{1}}$ takes the form
\be\n{b.40}
R_{\h{1}\h{1}}=-(\lap Q+U)\, .
\ee
The metric (\ref{b.19}) is a vacuum solution if the function $Q$ obeys
the equation
\be\n{b.41}
\lap Q=-U\, .
\ee

It is convenient to rewrite expression (\ref{b.29}) for $U$ in the
form
\be\n{b.42}
U=U_+ +U_-\, ,
\ee
where
\be\n{b.43}
U_+={1\over 2}(W_{i,j}+W_{j,i})(W_{\bar{i},\bar{j}}+
W_{\bar{j},\bar{i}})\, ,
\ee
\be\n{b.44}
U_-={1\over 2}(W_{\bar{i},j}+W_{j,\bar{i}})(W_{i,\bar{j}}+
W_{\bar{j},i})\, .
\ee
The calculations give
\be\n{b.45}
U_{\pm}={l^2 \over R^{4l+4}}(R^2P^2\pm I^2)\, ,
\ee
Here
\be\n{b.48}
P^2=2\sum_i p_i^2 \zeta^i\bar{\zeta}^i\, ,
\ee
\be\n{b.49}
I=2\sum_i p_i \zeta_i \bar{\zeta}^i\, .
\ee
Combining these relations one obtains
\be\n{b.52}
U={2l^2 P^2 \over R^{4l+2}}\, .
\ee

We write a solution $Q$ of the equation
\be\n{b.53}
\lap Q=-U\, ,
\ee
as a sum
\be\n{sol}
Q=\Phi+\Psi\, ,
\ee
where $\Psi$ is a special solution of the
inhomogeneous equation (\ref{b.53}) and $\Phi$ is an arbitrary solution
of the homogeneous equation
\be\n{b.54}
\lap \Phi =0\, .
\ee
Relation (\ref{lap1}) implies that
\be\n{b.55}
\Phi={\mu (u)\over R^{2(l-1)}}
\ee
with an arbitrary function $\mu (u)$.

To find a solution of the inhomogeneous equation
we use the ansatz
\be\n{b.56}
\Psi=a_l {P^2\over R^{4l}}+b_l {p^2\over R^{4l-2}}\, ,
\ee
where
\be
p^2=\sum_i p_i^2\, .
\ee
Using relations (\ref{lap1}) and (\ref{lap3}) one has
\be\n{b.57}
\lap \Psi={1\over R^{4l}} \left[ A_l {P^2\over R^{2}}+B_l p^2\right]\, ,
\ee
\be\n{b.58}
A_l=8l(l-1)a_l\hh B_l=4a_l+4l(2l-1)b_l\, .
\ee
Equations (\ref{b.52}) and (\ref{b.53}) give
\be\n{b.59}
a_l=-{l\over 4(l-1)}\hh
b_l={1\over 4(l-1)(2l-1)}\, .
\ee
Let us denote
\be\n{b.60}
{\cal B}=2\Psi+2\sum_i W_i \bar{W}_i\, .
\ee
Since
\be\n{b.61}
\sum_i W_i \bar{W}_i={P^2\over 2 R^{4l}}\, ,
\ee
one has
\be\n{b.62}
{\cal B}={1\over R^{4l-2}} \left[ \alpha_l {P^2\over R^{2}}+\beta_l
p^2\right]\, ,
\ee
where
\be\n{b.63}
\alpha_l={l-2\over 2(l-1)}\hh
\beta_l={1\over 2(l-1)(2l-1)}\, .
\ee

\subsection{Odd-dimensional space-time}

In the case of an odd dimensional space-time the calculations are
similar. We shall briefly give the main results omitting the details.

Let us denote $l=(D-3)/2$. As earlier we use 2 real coordinates,
$x^1=u$, $x^2=v$. We denote the other coordinates by $x^a$,
$a=3,\ldots,D$. They consist of $l$ sets of complex conjugated coordinates
$x^i=\zeta^i$, $x^{\bar{i}}=\bar{\zeta}^{i}$, where
$i=3,5, \ldots, 2l+1$ and $\bar{i}=i+1$, and one additional
real coordinate, which we denote by $z$, $x^{2l+3}=z$.

In this notations a partial
derivative $A_{,a}$ denotes the following set of
partial derivatives
\be\n{b.64}
A_{,i}=\pa_{\zeta^i} A\hh A_{,\bar{i}}=\pa_{\bar{\zeta}^i} A
\hh
A_{,z}=\pa_{z} A
\, .
\ee
We shall use the same convention for the indices connected with the
basic vectors and forms.
We denote
\be\n{b.65}
\eta^{ab}=\sum_{i}
[\delta^a_{i}\delta^b_{\bar{i}}+\delta^a_{\bar{i}}\delta^b_{i}]
+\delta^a_{z}\delta^b_{\bar{z}}\, .
\ee
Using this notation we can write the metric in the form
\be\n{b.66}
ds^2=-2\omega^{\h{1}}\omega^{\h{2}}+2\sum_i \w{i}\bw{i} + (\w{z})^2\, .
\ee
The basic forms $\omega^{\h{1}}$, $\omega^{\h{2}}$, $\w{i}$ and
$\bw{i}$ are given by (\ref{b.11})--(\ref{b.12}) with
$Q=Q(u,\zeta_i,\bar{\zeta}_i,z)$ and
$W^i=W^i(u,\zeta_i,\bar{\zeta}_i,z)$ and
\be\n{b.67}
\omega^{\h{z}}=dz\, .
\ee
The metric is
\[
ds^2=- du dv +2\sum_i d\zeta^i d\bar{\zeta}^i+dz^2
\, .
\]
\be\n{b.68}
+ 2\sum_i (W_i d\zeta^i+W_{\bar{i}} d\bar{\zeta}^i) du
+2(Q+\sum_i W_i W_{\bar{i}})du^2 \, .
\ee
It is convenient to denote $W_a$ an object which besides the
components $W_i$ and $W_{\hat{i}}$ has one more additional component
$W_z=0$. Using these notations it is possible to show that the
non-vanishing components of $\lambda_{\h{\mu}\h{\nu}\h{\lambda}}$ and
$\gamma_{\h{\mu}\h{\nu}\h{\lambda}}$ are given by relations
(\ref{b.20})--(\ref{b.24}). For the non-vanishing components of the
Ricci tensor one has
\be\n{b.69}
R_{\h{1}\h{1}}=-\lap Q+ Z_{,u}-{\cal D} Z-U\, ,
\ee
\be\n{b.70}
R_{\h{1}{a}}=-{1\over 2} \lap W_{a}+{1\over 2}Z_{,a}\, .
\ee
Here
\be\n{b.70a}
{\cal D}=\sum_i ({W}^{i} \pa_{\z^i}+\bar{W}^{i} \pa_{\bz^i})\, ,
\ee
\be\n{b.71}
\lap=\eta^{ab}\pa_a\pa_b=2\sum_i
\pa_{\zeta^i}\pa_{\bar{\zeta}^i}+\pa^2_z\, ,
\ee
\be\n{b.72}
Z=\eta^{ab}W_{a,b}=\sum_i (W_{i,\bar{i}}+W_{\bar{i},i})\, ,
\ee
\be\n{b.73}
U={1\over 2}(\eta^{ac}\eta^{bd}+\eta^{ad}\eta^{bc}) W_{a,b}W_{c,d}\, .
\ee

To obtain a solution we choose
\be\n{b.74}
W_i=-i p_i{\bar{\zeta}^i\over R^{(2l+1)}}\, ,
\ee
\be\n{b.75}
R^2=2\sum_i \zeta^i \bar{\zeta}^i+z^2\, .
\ee
Here $\mu(u)$ and $p_i=p_i(u)$ are arbitrary functions of $u$.
We demonstrate now that for this choice the components
$R_{\h{1}\h{a}}$  of Ricci tensor vanish.

It is easy to check that
\be\n{b.76}
W_{i,j}=i(2l+1) p_i {\bar{\zeta}^i \bar{\zeta}^j\over R^{(2l+3)}}\, ,
\ee
\be\n{b.77}
W_{\bar{i},\bar{j}}=-i(2l+1) p_i {{\zeta}^i {\zeta}^j\over R^{(2l+3)}}\, ,
\ee
\be\n{b.78}
W_{i,\bar{j}}={i \over R^{(2l+1)}}\left[ (2l+1) p_i{\bar{\zeta}^i
{\zeta}^j\over R^{2}}-p_i \delta_{ij}
\right]\, ,
\ee
\be\n{b.79}
W_{\bar{i},{j}}=-{i \over R^{(2l+1)}}\left[ (2l+1) p_i{{\zeta}^i
\bar{\zeta}^j\over R^{2}}-p_i \delta_{ij}
\right]\, .
\ee
\be\n{b.80}
W_{i,z}=i(2l+1) p_i {\bar{\zeta}^i z\over R^{(2l+3)}}\, ,
\ee
\be\n{b.81}
W_{\bar{i},z}=-i(2l+1) p_i {\zeta^i z\over R^{(2l+3)}}\, .
\ee

It is easy to see that for each $i$, $W_{i,\bar{i}}+W_{\bar{i},i}=0$,
and hence $Z=0$.
We also have (outside a singular point $R=0$)
\be\n{lap1a}
\lap \left({1\over R^{2m}}\right)={2m(2m-2l+1)\over R^{2(m+1)}}\, ,
\ee
\be\n{lap2a}
\lap \left({\zeta^i\over R^{2m}}\right)={2m(2m-2l-1)\zeta^i\over R^{2(m+1)}}\, ,
\ee
\be\n{lap3a}
\lap \left({\zeta^i\bar{\zeta}^i\over R^{2m}}\right)={2\over R^{2m}}\left[
1+{m(2m-2l-3)\zeta^i\bar{\zeta}^i\over R^2}\right]\, ,
\ee
\be\n{lap4a}
\lap \left({z^2\over R^{2m}}\right)={2\over R^{2m}}\left[
1+{m(2m-2l-3)z^2\over R^2}\right]\, .
\ee
The equation (\ref{lap2a}) implies
\be\n{b.85}
\lap W_i=\lap W_{\bar{i}}=0\, .
\ee
Thus $R_{\h{1}\h{a}}=0$ and $R_{\h{1}\h{1}}$ takes the form
\be\n{b.86}
R_{\h{1}\h{1}}=-(\lap Q+U)\, .
\ee
The metric (\ref{b.68}) is a vacuum solution if the function $Q$ obeys
the equation
\be\n{b.87}
\lap Q=-U\, .
\ee

The function $U$ defined by (\ref{b.29}) can be rewritten as
\be
U=U_+ +U_- +U_0\, ,
\ee
where $U_{\pm}$ are defined by (\ref{b.43}) and (\ref{b.44}) and
\be
U_0=\sum_i W_{i,z} W_{\bar{i},z}\, .
\ee
The calculations give
\be
U_{\pm}={(2l+1)^2\over 4 R^{4l+6}}[ P^2(R^2-z^2)\pm I^2]\, ,
\ee
\be
U_0={(2l+1)^2\over 2 R^{4l+6}} P^2 z^2\, .
\ee
Thus one has
\be\n{UU}
U={(2l+1)^2  P^2 \over 2 R^{4l+4}}\, .
\ee

We write a solution $Q$ of (\ref{b.87}) in the form (\ref{sol}),
$Q=\Phi+\Psi$, where as earlier $\Phi$ is a solution of the
homogeneous equation and $\Psi$ is a special solution of the
inhomogeneous one. Relation (\ref{lap1a}) implies that
\be
\Phi={\mu(u)\over R^{2l-1}}\, ,
\ee
where $\mu(u)$ is an arbitrary function of $u$.

To find a solution of the inhomogeneous equation
we use the ansatz \cite{anz}
\be\n{b.100}
\Psi=a_l {P^2\over R^{4l+2}}+b_l {p^2\over R^{4l}}\, .
\ee
Using relations (\ref{lap1}) and (\ref{lap3}) one has
\be\n{b.101}
\lap \Psi={1\over R^{4l+2}} \left[ A_l {P^2\over R^{2}}+B_l p^2\right]\, ,
\ee
\be\n{b.102}
A_l=2(4l^2-1)a_l\hh B_l=4a_l+4l(2l+1)b_l\, .
\ee
Equations (\ref{b.87}) and (\ref{UU}) give
\be\n{b.103}
a_l=-{(2l+1)\over 4(2l-1)}\hh
b_l={1\over 4l(2l-1)}\, .
\ee
Let us denote
\be\n{b.104}
{\cal B}=2\Psi+2\sum_i W_i \bar{W}_i\, .
\ee
Since
\be\n{b.105}
\sum_i W_i \bar{W}_i={P^2\over 2 R^{4l+2}}\, ,
\ee
one has
\be\n{b.106}
{\cal B}={1\over R^{4l}} \left[ \alpha_l {P^2\over R^{2}}+\beta_l
p^2\right]\, ,
\ee
where
\be\n{b.107a}
\alpha_l={2l-3\over 2(2l-1)}\hh
\beta_l={1\over 2l(2l-1)}\, .
\ee


\begin{thebibliography}{}
\bibitem{Noether} Tensor $M^{\sigma\rho}_\nu$
in (\ref{3.4}) differes from
the angular momentum tensor derived with the help of the Noether
theorem by the term $\partial^\lambda Q^{\sigma\rho}_{\lambda\nu}$,
where  $Q^{\sigma\rho}_{\lambda\nu}=
F_{\lambda\nu}(A^\sigma x^\rho-A^\rho x^\sigma)$. This addition
does not contribute to the integral of the angular momentum.
\bibitem{zz} To simplify expressions, in this section we use notations
slightly different from those adopted in appendix~B. Namely, the index
$i$ takes values $i=1,\ldots,l$. We also use complex coordinates $z^i$
related to $\zeta^i$ as follows $z^i=\zeta^{2i+1}$ and
$\bar{z}^i=\bar{\zeta}^{2i+1}=\zeta^{2i+2}$.
\bibitem{anz} In order to make the ansatz more general, one may add
to the expression for $\Psi$ an additional term of the form $c_l
z^2/R^{4l+2}$. Using (\ref{lap4a}) one can see that a similar term
would appear in $\lap \Psi$. The other terms in $\Psi$ do not
generate such a term, and $U$ does not contain it. Thus it is
possible to conclude that $c_l=0$.
\end{thebibliography}
\end{document}